\documentstyle[a4wide,epsfig,11pt,titlepage]{article}

\newcounter{nref}
\newcommand{\bbib}{%
  \renewcommand{\refname}{\large\bf References}%
  \begin{thebibliography}{99}%
  \setcounter{enumiv}{\thenref}}
\newcommand{\ebib}{%
  \end{thebibliography}%
  \setcounter{nref}{\arabic{enumiv}}}
\newcommand{\head}[3]{%
  \setcounter{nref}{0}%
  \thispagestyle{empty}%
  \section*{\LARGE\bf #1}%
  \stepcounter{section}%
  \addcontentsline{toc}{section}{#1}%
  \large\itshape%
  #2\\\vspace{0.1pt}\\%
  #3%
  \normalsize\upshape%
  \bigskip}

\def \la {\mathrel{\vcenter
     {\offinterlineskip \hbox{$<$}\hbox{$\sim$}}}}

\begin{document}


\head{Core-Collapse Supernovae: Modeling between\\ Pragmatism and 
      Perfectionism}
     {H.-Th.\ Janka, R.\ Buras, F.S.\ Kitaura Joyanes, A.\ Marek, 
      and M.\ Rampp}
     {Max-Planck-Institut f\"ur Astrophysik, Karl-Schwarzschild-Str.\ 1,
      D-85741 Garching, Germany}

\subsection*{Abstract}
We briefly summarize recent efforts in Garching for modeling
stellar core collapse and post-bounce evolution in one and two
dimensions. The transport of neutrinos of all flavors is treated
by iteratively solving the coupled system of frequency-dependent
moment equations together with a model Boltzmann equation which 
provides the closure. A variety of progenitor stars, different
nuclear equations of state, stellar rotation, and global asymmetries
due to large-mode hydrodynamic instabilities have been investigated 
to ascertain the road to finally successful, convectively supported
neutrino-driven explosions.

\setcounter{section}{0}

\section{Methods}
\label{janka.sec_methods}

Our neutrino-hydrodynamic simulations were performed
with a conservative, high-resolution shock-capturing 
scheme that employs an exact Riemann solver and integrates the 
hydrodynamics equations with third order accuracy in space and
second order accuracy in time. This code is coupled to a
neutrino transport scheme that solves the energy-dependent
moment equations of neutrino number, energy, and momentum
of order $(v/c)$
by making use of a variable Eddington factor closure which is
calculated from a model Boltzmann equation~\cite{janka.ref1}.
In addition to radial derivatives, the moment equations 
are supplemented by ($v_{\theta}$-dependent)
terms that correspond to the advection of neutrinos
with fluid motions in lateral direction (with 
$v_{\theta}$ being the lateral component of the velocity 
vector in spherical coordinates),
and the equation of motion of the stellar plasma contains
also lateral gradients of the neutrino 
pressure~\cite{janka.ref2}.
General relativistic effects are taken into account 
approximately by (i) a generalized gravitational potential 
(the radial part of which is constructed by a comparison
with the Tolman-Oppenheimer-Volkoff equation, including
terms due to fluid motion and neutrino effects) and (ii) 
relativistic redshift terms
in the transport equations~\cite{janka.ref1}. A comparison
with a fully relativistic treatment in spherical symmetry
yields very satisfactory results~\cite{janka.ref3}.
We have continuously modernized our treatment of neutrino
interactions in the supernova medium, now including a 
variety of reactions and improvements (e.g., nucleon-nucleon
bremsstrahlung; interactions between neutrinos of different
flavors; correlations, weak magnetism and 
recoil in neutrino-nucleon interactions; electron captures
on heavy nuclei according to recent calculations by Langanke
and coworkers) in extension to the standard treatment of
neutrino-matter interactions in supernova simulations (see
Refs.~\cite{janka.ref1,janka.ref2,janka.ref4} and the references 
to a long list of original papers therein).

\begin{figure}[htp!]
  \centerline{\epsfxsize=0.38\textwidth \epsfclipon \epsffile{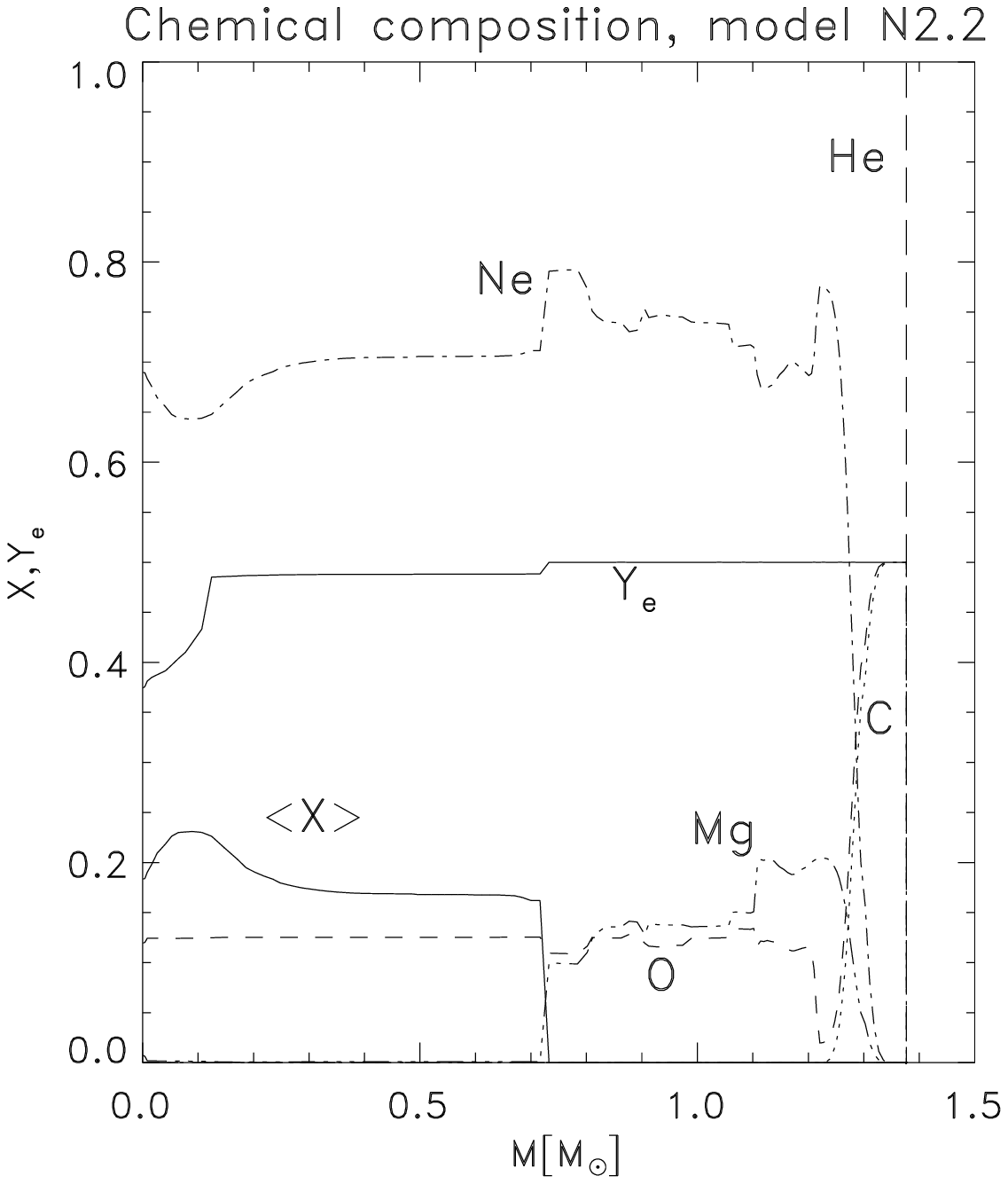}
              \epsfxsize=0.60\textwidth \epsfclipon \epsffile{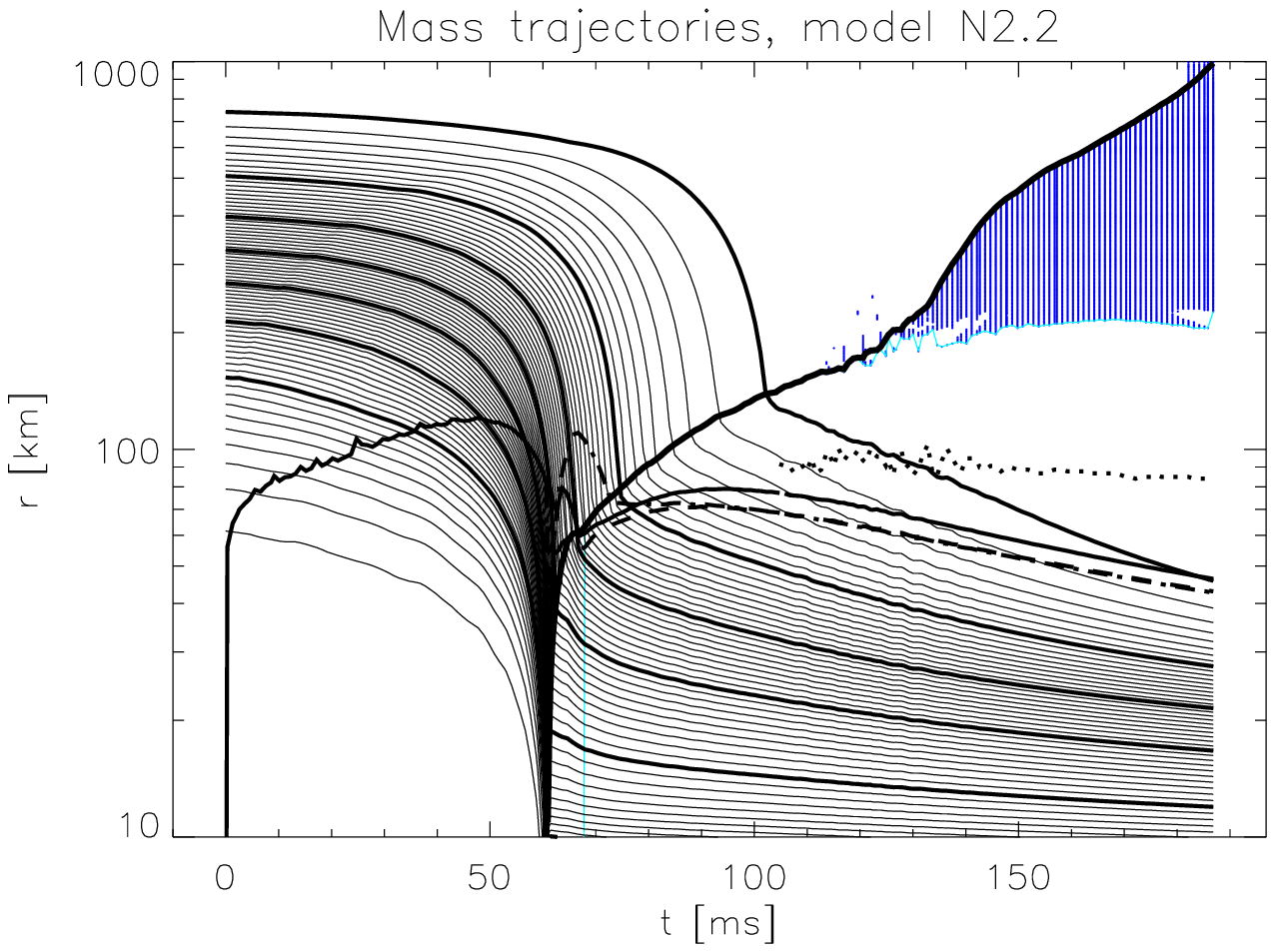}}
  \caption{\small 
{\em Left:} Composition in the highly degenerate 1.38$\,M_{\odot}$ 
core of an 8--10$\,M_{\odot}$ progenitor~\cite{janka.ref_nom87} 
with O, Ne and Mg in the central region enclosed by an C-O shell.
$\left\langle X \right\rangle$ denotes the mass fraction of a 
representative neutron-rich heavy nucleus that appears in 
nuclear statistical equilibrium, and $Y_e$ is the electron
fraction.
{\em Right:} Mass trajectories, shock position (thick solid line 
that rises to the upper right corner of the plot, where it reaches
the surface of the C-O shell), and neutrinospheres ($\nu_e$: solid;
$\bar\nu_e$: dash-dotted; $\nu_{\mu},\,\bar\nu_{\mu},\,\nu_{\tau},\,
\bar\nu_{\tau}$: dashed), and gain radius (dotted) for the collapsing
O-Ne-Mg core. The mass trajectories are equidistantly spaced with
intervals of 0.01$\,M_{\odot}$. The (blue) hatched region is 
characterized by a dominant mass fraction of He. (The plots were
taken from Ref.~\cite{janka.ref_kit03}.)
}
  \label{janka.fig1}
\end{figure}

\begin{figure}[t!]
  \centerline{\epsfxsize=0.49\textwidth \epsfclipon \epsffile{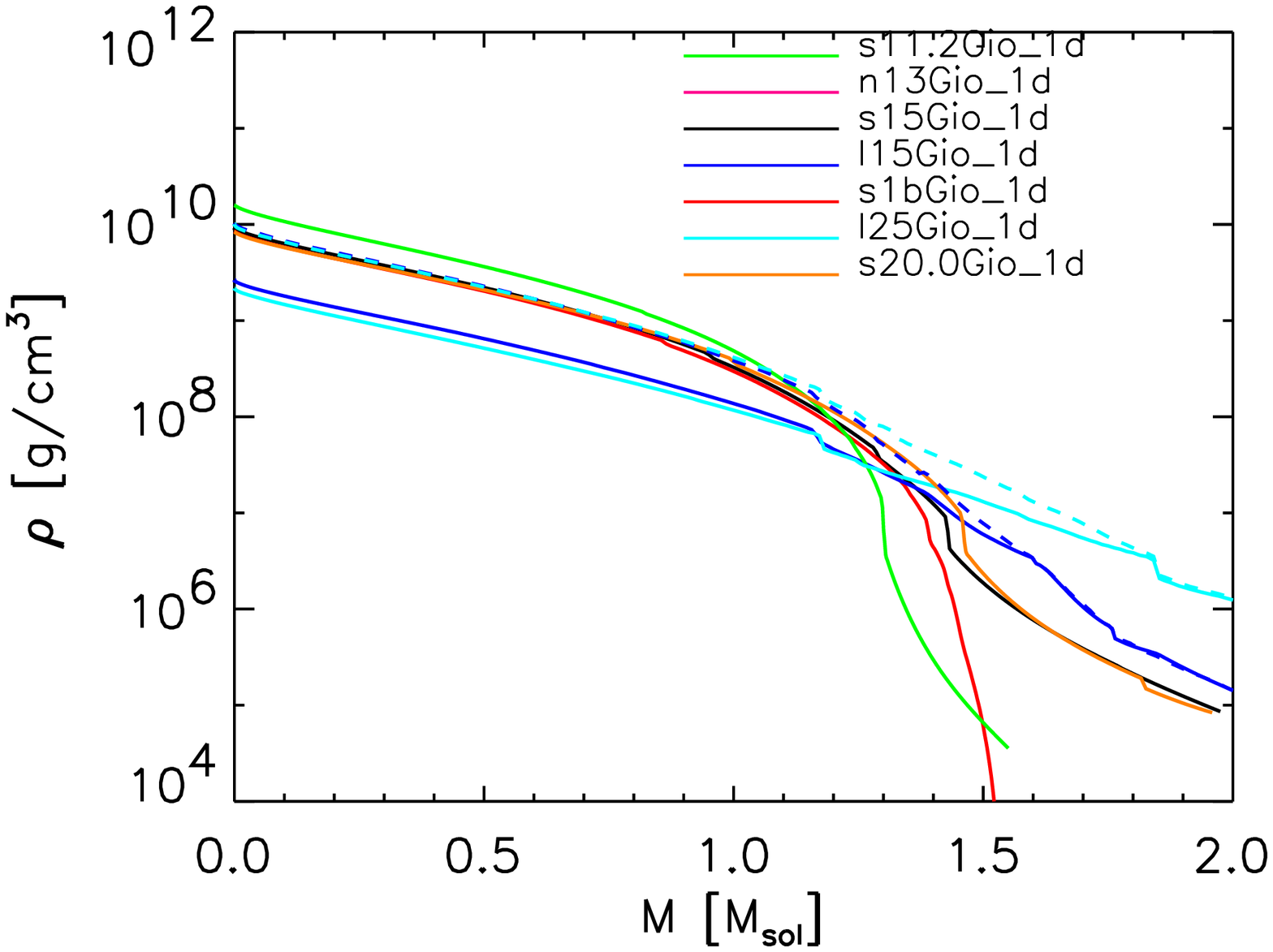}
              \epsfxsize=0.49\textwidth \epsfclipon \epsffile{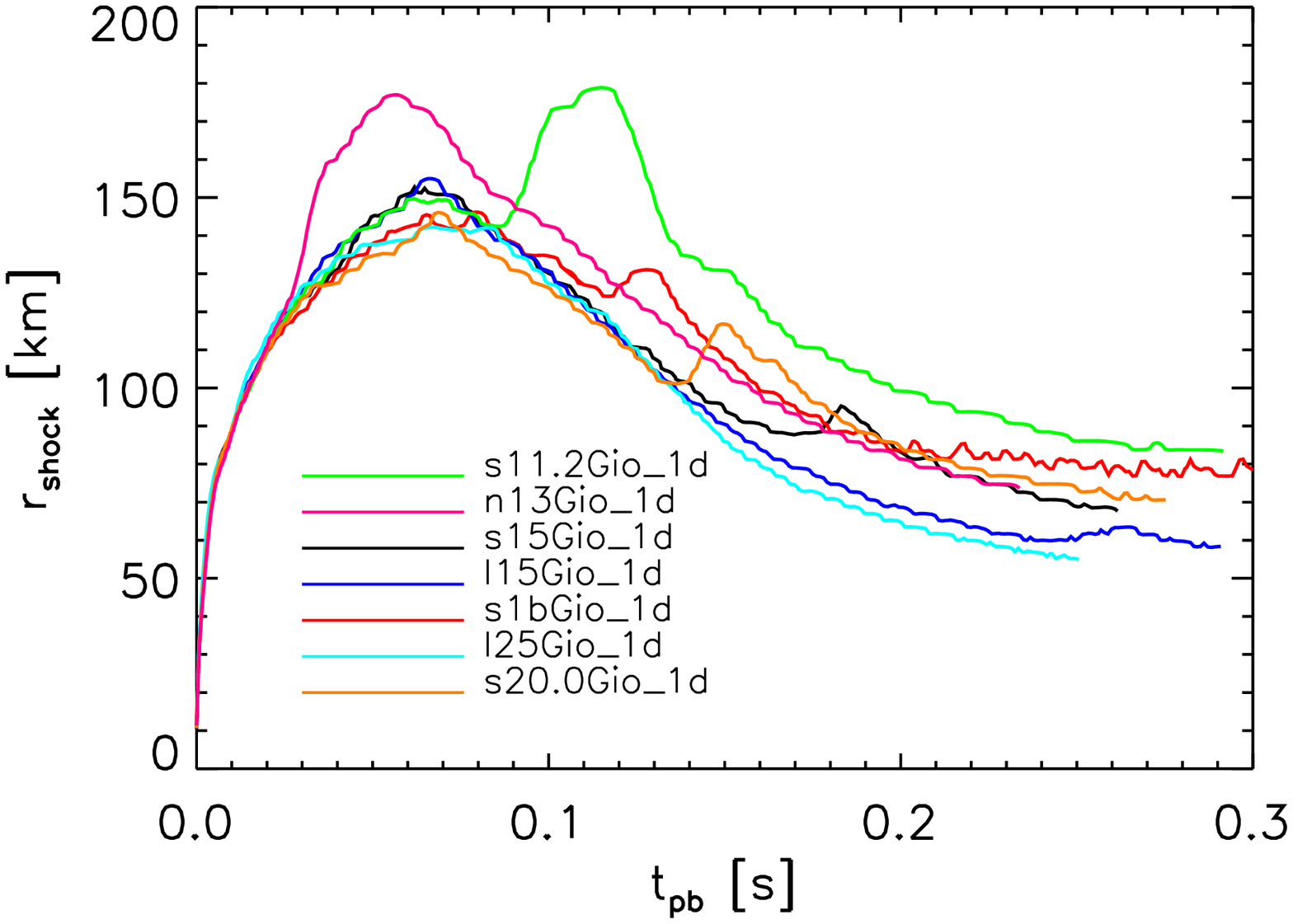}}
  \centerline{\epsfxsize=0.49\textwidth \epsfclipon \epsffile{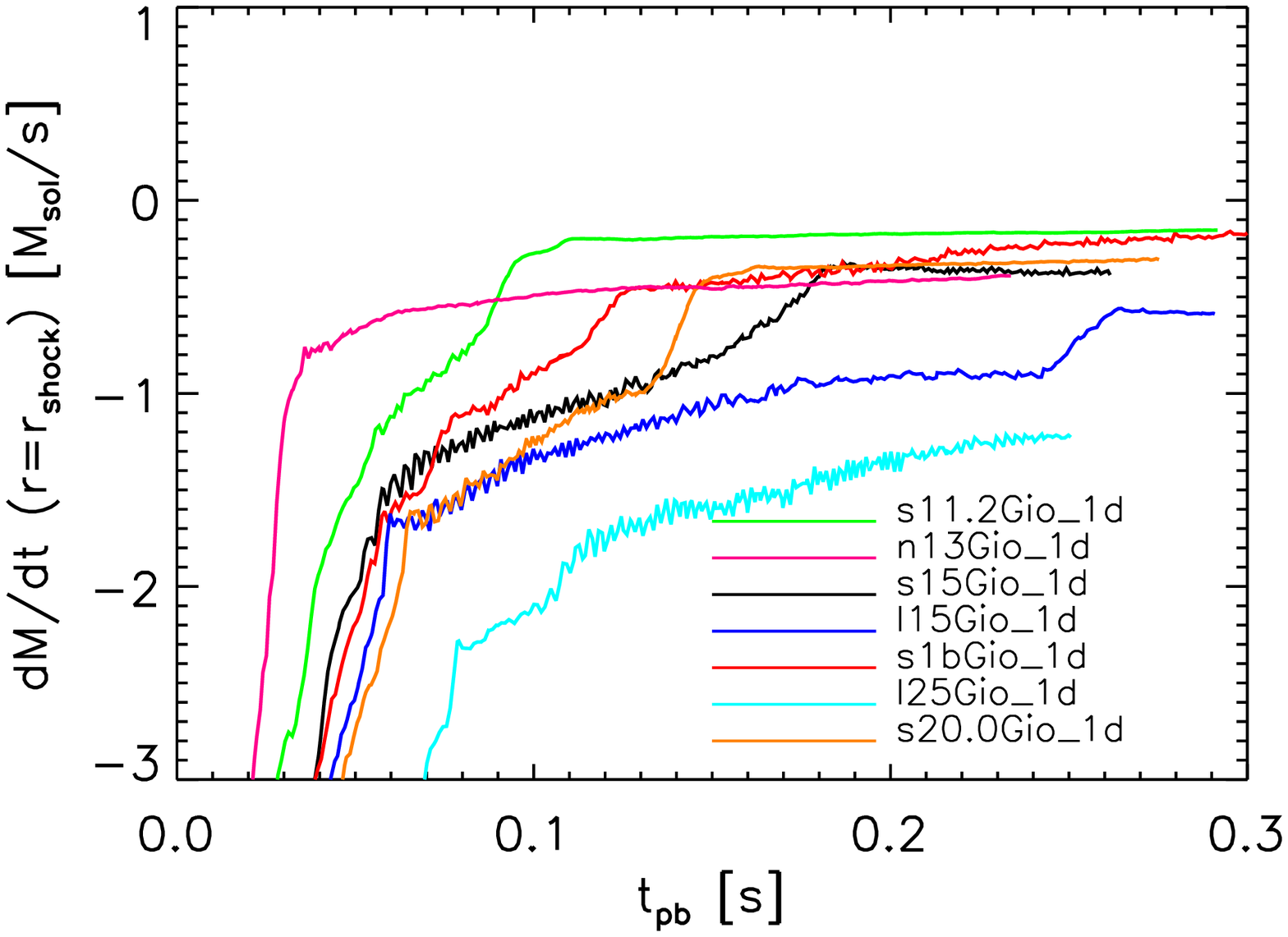}
              \epsfxsize=0.49\textwidth \epsfclipon \epsffile{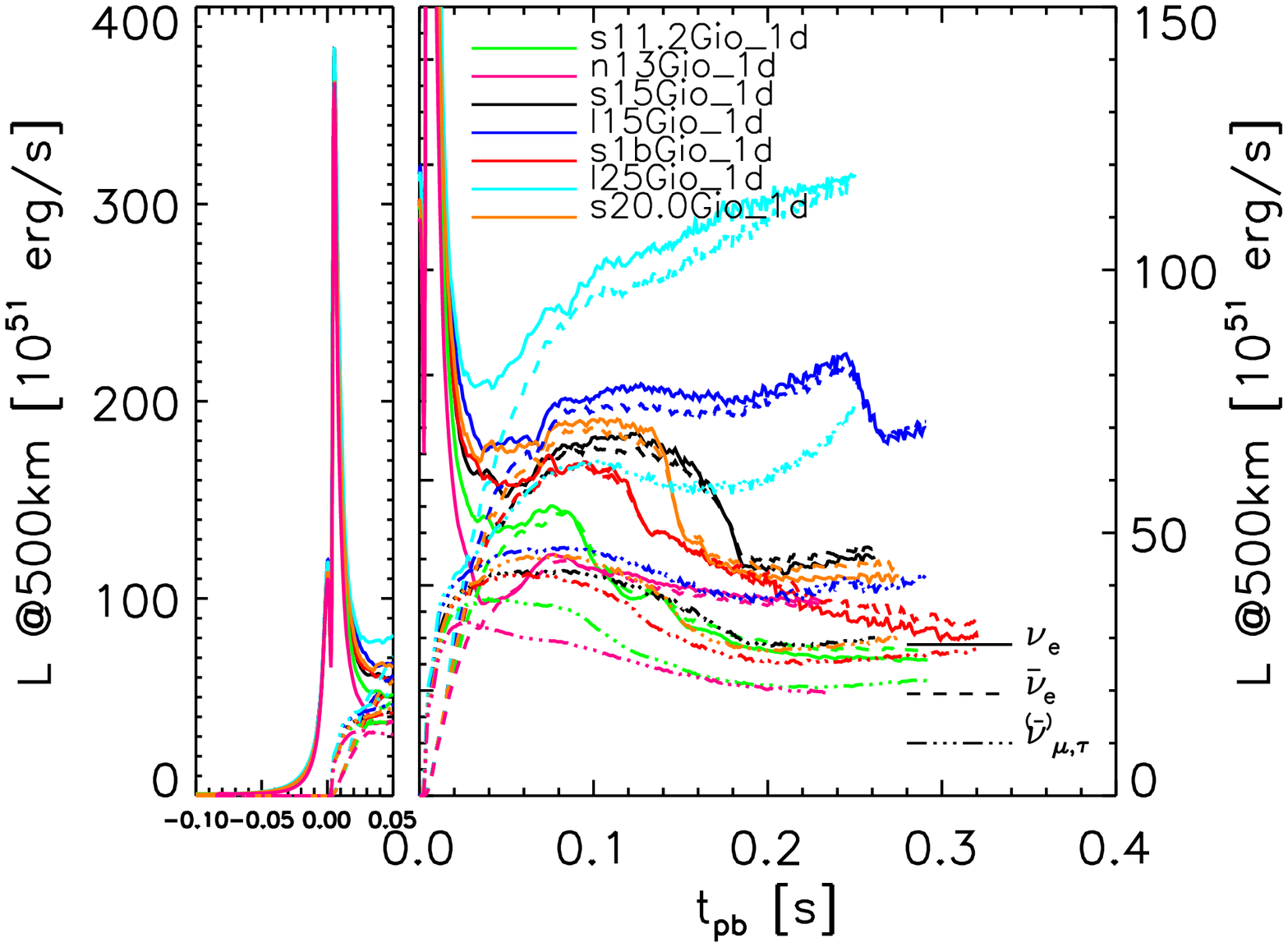}}
  \caption{\small 
{\em Top left:} Density profiles of a sample of progenitor cores
for stars with main sequence masses between 11.2$\,M_{\odot}$ and 
25$\,M_{\odot}$ according to Ref.~\cite{janka.ref_nomhas88} (13$\,M_{\odot}$,
n13Gio\_1d), Ref.~\cite{janka.ref_wooheg02} (11.2, 15, 20$\,M_{\odot}$,
s11.2Gio\_1d, s15Gio\_1d, s20.0Gio\_1d, respectively),
Ref.~\cite{janka.ref_woowea95}
(a Type Ib progenitor, s1bGio\_1d), and Ref.~\cite{janka.ref_limetal00}
(15 and 25$\,M_{\odot}$, l15Gio\_1d and l25Gio\_1d, respectively).
The dashed lines show profiles of the latter models after we have
evolved them to a central density similar to that of the other
progenitors.
{\em Top right:} Shock positions vs.\ time for the set of progenitors.
No explosions were obtained in spherical symmetry.
{\em Bottom left:} Mass accretion rates by the shock as functions of
time. The steep density gradients at composition interfaces cause
sudden drops of $|\dot M|$.
{\em Bottom right:} Luminosities of $\nu_e$ (solid lines), $\bar\nu_e$
(dashed) and heavy-lepton $\nu$'s and $\bar\nu$'s (dash-dotted),
measured by a comoving observer at 500$\,$km,
around the prompt $\nu_e$ burst (left panel) and during the
post-bounce evolution (right panel).
}
  \label{janka.fig2}
\end{figure}

\section{Results}

In the past two years we have applied our new neutrino-hydrodynamics code
to core-collapse simulations in one and two dimensions for a broad 
variety of progenitors,
spanning the range from O-Ne-Mg cores on the low-mass end
($\sim$8--10$\,M_{\odot}$) to $25\,M_{\odot}$ on the 
high-mass side.
We have investigated different input for neutrino-matter
interactions and have recently also tested the effects of different
nuclear equations of state. This work was mostly done in course of the 
PhD Thesis of R.~Buras and the Diploma Theses of 
F.S.~Kitaura Joyanes and A.~Marek.
Below some main results are briefly summarized.

\subsection{Explosion of an O-Ne-Mg core}

The main improvement of our new simulations of O-Ne-Mg core collapse
--- which we did so far only in spherical symmetry --- 
compared to previous approaches is the
more accurate treatment of neutrino transport and neutrino-matter
interactions. Using the nuclear equation of state (EoS) of 
Ref.~\cite{janka.ref_latswe91}, we could not 
confirm~\cite{janka.ref_kit03} that
a prompt explosion occurs as found in calculations with simpler
neutrino treatment but also with a 
different EoS~\cite{janka.ref_hiletal84}. The shock is created
at a mass coordinate of $\sim$0.45$\,M_{\odot}$ 
(defined by the moment when the postshock entropy first exceeds
3$\,k_{\mathrm B}$ per nucleon) and stalls
(defined by the time when the postshock velocity becomes negative)
only 1.2$\,$ms later at $\sim$0.8$\,M_{\odot}$, still
well inside the neutrinosphere and {\em before} energy losses by
the prompt $\nu_e$ burst could have contributed to its damping. 
We also do not find the powerful shock revival by neutrino heating
as seen in Ref.~\cite{janka.ref_maywil88}
(cf.\ also Ref.~\cite{janka.ref_baretal87}). Instead, the shock
continuously expands due to the monotonically decreasing preshock 
density associated with the steeply dropping density at the 
interface between C-O shell and He shell (Fig.~\ref{janka.fig1}). 
Towards the end of 
our simulation the mass accretion rate by the shock has 
correspondingly dropped to less than 0.03$\,M_{\odot}\,$s$^{-1}$.
Although our simulation is not yet finally conclusive in this
point, we see indications that a neutrino-driven wind begins
to fill the volume between neutron star surface and shock and
will lead to mass ejection with a rather low ``explosion''
energy and little Ni production, very similar to the outcome
of simulations of accretion induced white dwarf collapse to
neutron stars~\cite{janka.ref_woobar92}. But the results of
O-Ne-Mg core collapse can be sensitive to the properties of the
nuclear EoS~\cite{janka.ref_fry99}. We plan to investigate
this in future simulations.

\subsection{One-dimensional core collapse for progenitors
               between 11$\,M_{\odot}$ and 25$\,M_{\odot}$}

In spherical symmetry none of the investigated core collapse models, 
all evolved with the EoS of Lattimer and 
Swesty~\cite{janka.ref_latswe91}, developed
an explosion until we stopped the simulations at typically
300$\,$ms after bounce. This confirms the results of fully relativistic
simulations by Liebend\"orfer et al.~\cite{janka.ref_lieetal04}.
In spite of a rapidly dropping rate of mass accretion through the 
shock (Fig.~\ref{janka.fig2}) neutrino heating is not strong enough
to reverse the postshock infall of the accreted matter, and
the shock retreats in all cases, following the contraction of
the neutrinosphere, after it had temporarily expanded 
to a maximum radius between 140$\,$km and 
180$\,$km (Fig.~\ref{janka.fig2}).
The latter expansion was driven by the accumulation of accreted 
matter between shock and neutrinosphere during the early 
post-bounce phase when the accretion rate is very high.
The bumps visible in the shock trajectories (Fig.~\ref{janka.fig2})
correspond to the density and entropy discontinuities at the 
interfaces between shells of different composition, where the shock
experiences a sudden drop of the mass accretion rate.
Note the astonishing similarity of the prompt $\nu_e$ burst, which 
is a consequence of the very similar structure of the inner
$\sim$1$\,M_{\odot}$ of the Fe core for different progenitors
before collapse (Fig.~\ref{janka.fig2}) and the fact that the
structural similarity is still present at bounce.
The large luminosity differences during the 
post-bounce evolution are caused by model-to-model
variations of the structure near and beyond the outer 
edge of the iron core (cf. also~\cite{janka.ref_lieetal03}).
First core-collapse simulations of 15$\,M_{\odot}$ and 
25$\,M_{\odot}$ progenitors with electron captures on heavy
nuclei being treated according to Ref.~\cite{janka.ref_lang03} 
suggest that these findings also apply when 
electron captures on nuclei instead of free protons dominate
during the infall phase.

\begin{figure}[t!]
  \centerline{\epsfxsize=0.34\textwidth \epsfclipon \epsffile{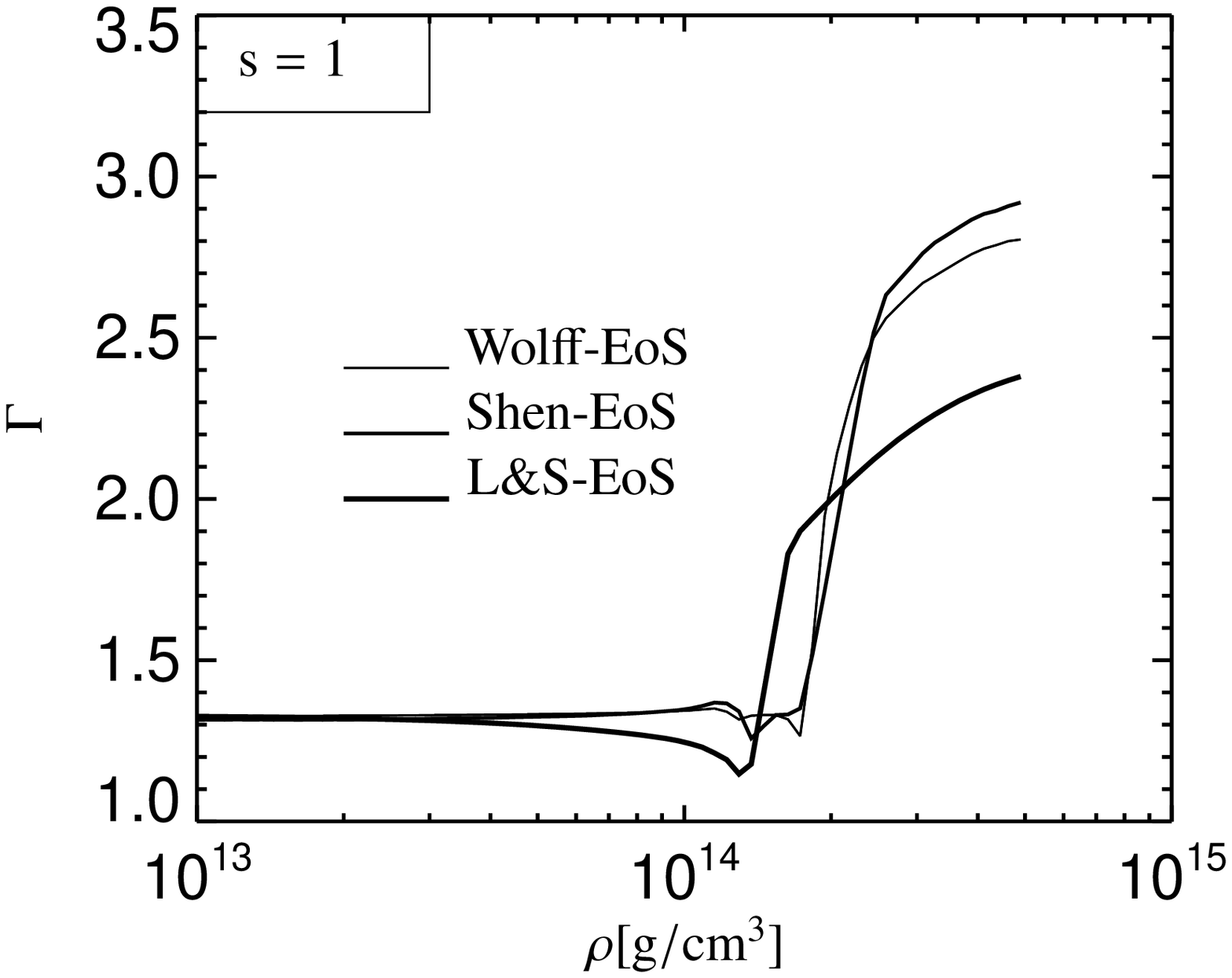}
              \epsfxsize=0.30\textwidth \epsfclipon \epsffile{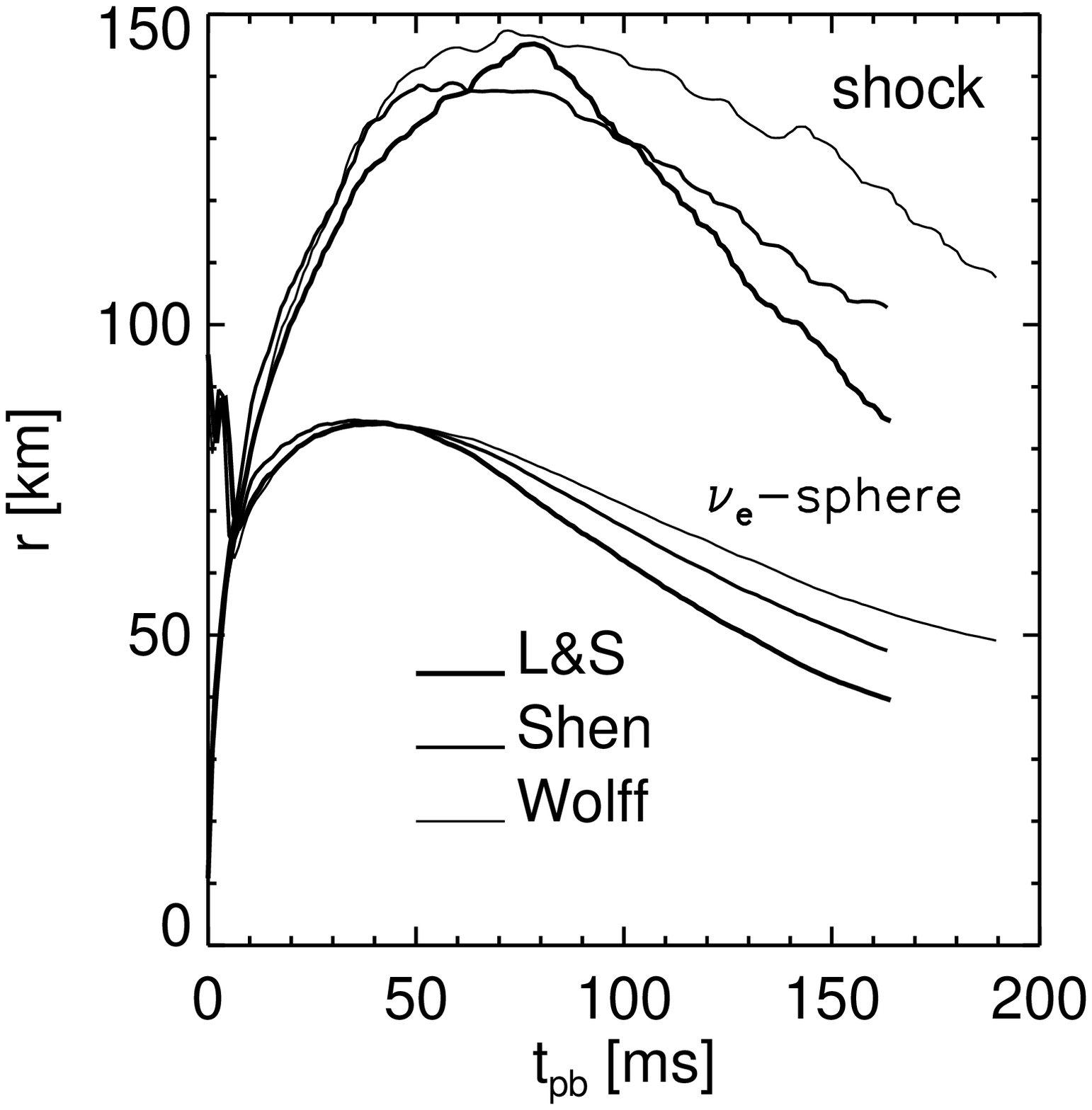}
              \epsfxsize=0.35\textwidth \epsfclipon \epsffile{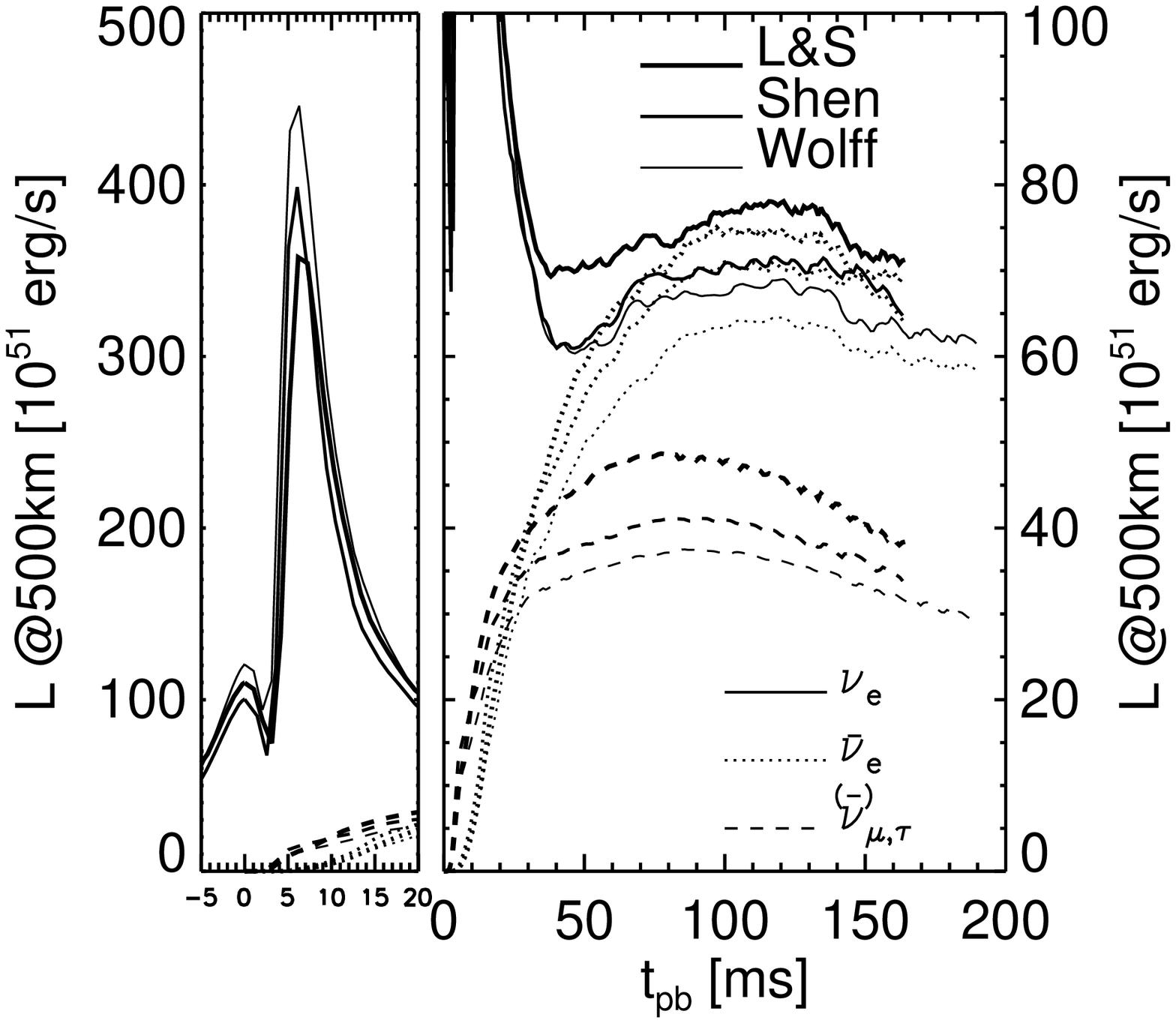}}
  \caption{\small 
{\em Left:} Adiabatic index 
$\Gamma \equiv \left ({\mathrm d}\ln P/{\mathrm d}\ln\rho\right )_s$ 
for an entropy $s = 1\,k_{\mathrm B}$ per nucleon (and an electron
fraction $Y_e = 0.4$) for the equations of
state of Ref.~\cite{janka.ref_hiletal84} (``Wolff'', thin lines),
Ref.~\cite{janka.ref_shenetal98} (``Shen'', medium) 
and Ref.~\cite{janka.ref_latswe91} (``L\&S'', thick).
{\em Middle:} Shock positions and neutrinospheric radii of $\nu_e$
as functions of time for collapse simulations of a 15$\,M_{\odot}$
progenitor (Model~s15a28~\cite{janka.ref_hegwww}) 
with the three nuclear equations of state.
{\em Right:} Prompt $\nu_e$ burst (left panel) and post-bounce
luminosities of $\nu_e$ (solid lines), $\bar\nu_e$ (dotted) and
heavy-lepton $\nu$'s and $\bar\nu$'s (dashed) for the core collapse
simulations of the 15$\,M_{\odot}$ star with the three different 
equations of state. (The plots were taken from Ref.~\cite{janka.ref_mar03}.)
}
  \label{janka.fig3}
\end{figure}

\subsection{Variations of the nuclear equation of state}

For a 15$\,M_{\odot}$ progenitor 
(Model~s15a28~\cite{janka.ref_hegwww}) we have investigated
the influence of the nuclear EoS on core collapse, shock formation,
and post-bounce evolution in three different cases. We ran simulations
with\\
\noindent
(i) our standard EoS (Lattimer and Swesty, 
``L\&S'')~\cite{janka.ref_latswe91}, which is based on a 
compressible liquid drop model and employs a Skyrme force for the
nucleon interaction (our choice of the
compressibility modulus of bulk nuclear matter is 180$\,$MeV, 
and the symmetry energy parameter 29.3$\,$MeV,
but the differences in the supernova evolution caused by other 
values of the compressibility were shown to be 
minor~\cite{janka.ref_thompetal03,janka.ref_sweetal94});\\
\noindent
(ii) a new relativistic mean field EoS (Shen et al.,
``Shen'')~\cite{janka.ref_shenetal98}
with a compressibility of nuclear matter of 281$\,$MeV and a
symmetry energy of 36.9$\,$MeV;\\
\noindent
(iii) an EoS that was constructed by Hartree-Fock calculations
with a Skyrme force for the nucleon-nucleon 
interaction (Wolff and Hillebrandt, 
``Wolff'')~\cite{janka.ref_hiletal84} and
has a compressibility of 263$\,$MeV and a symmetry energy of
32.9$\,$MeV.\\  
The three equations of state show, for example, clear differences
in the adiabatic index 
$\Gamma\equiv \left ({\mathrm d}\ln P/{\mathrm d}\ln\rho\right )_s$ 
around and above the phase transition to homogeneous nuclear 
matter (Fig.~\ref{janka.fig3}).

\vspace{0.2truecm}
\noindent
Significant differences were also found during the core-collapse
calculations:\\
\noindent
(a) in the nuclear composition of the progenitor core and during
infall;\\
\noindent
(b) in the maximum density reached at bounce, and the density in
hydrostatic equilibrium afterwards;\\
\noindent
(c) in the shock formation point, which is shifted outward by
$\sim$0.05$\,M_{\odot}$ in case of the stiffer Wolff and Shen
EoSs~\cite{janka.ref_shenetal98,janka.ref_hiletal84};\\
\noindent
(d) in the shock stagnation point (defined by the time when the
postshock velocities start to become negative), which moves out
by up to $\sim$0.1$\,M_{\odot}$ (or $\sim$10$\,$km) for the 
stiffer EoSs;\\
\noindent
(e) concerning the existence and duration of a 
post-bounce phase where a shell with a sizable mass fraction of
heavy nuclei occurs transiently outside of the nuclear core;\\
\noindent
(f) in the radius of maximum shock expansion and subsequent 
shock contraction, which closely follows the contraction behavior
of the nascent neutron star (Fig.~\ref{janka.fig3});\\
\noindent
(g) in the peak luminosity during the prompt $\nu_e$ burst and the 
evolution of the post-bounce neutrino luminosities 
(Fig.~\ref{janka.fig3});\\
\noindent
(h) despite of differences in details, an application of the 
Ledoux criterion suggests a qualitatively similar behavior
of convectively unstable regions in the neutron star and 
in the neutrino-heating layer.\\
The publication of a detailed discussion of these results is 
in preparation (Marek et al.\ 2004).

\begin{figure}[t!]
  \centerline{\epsfxsize=0.49\textwidth \epsfclipon \epsffile{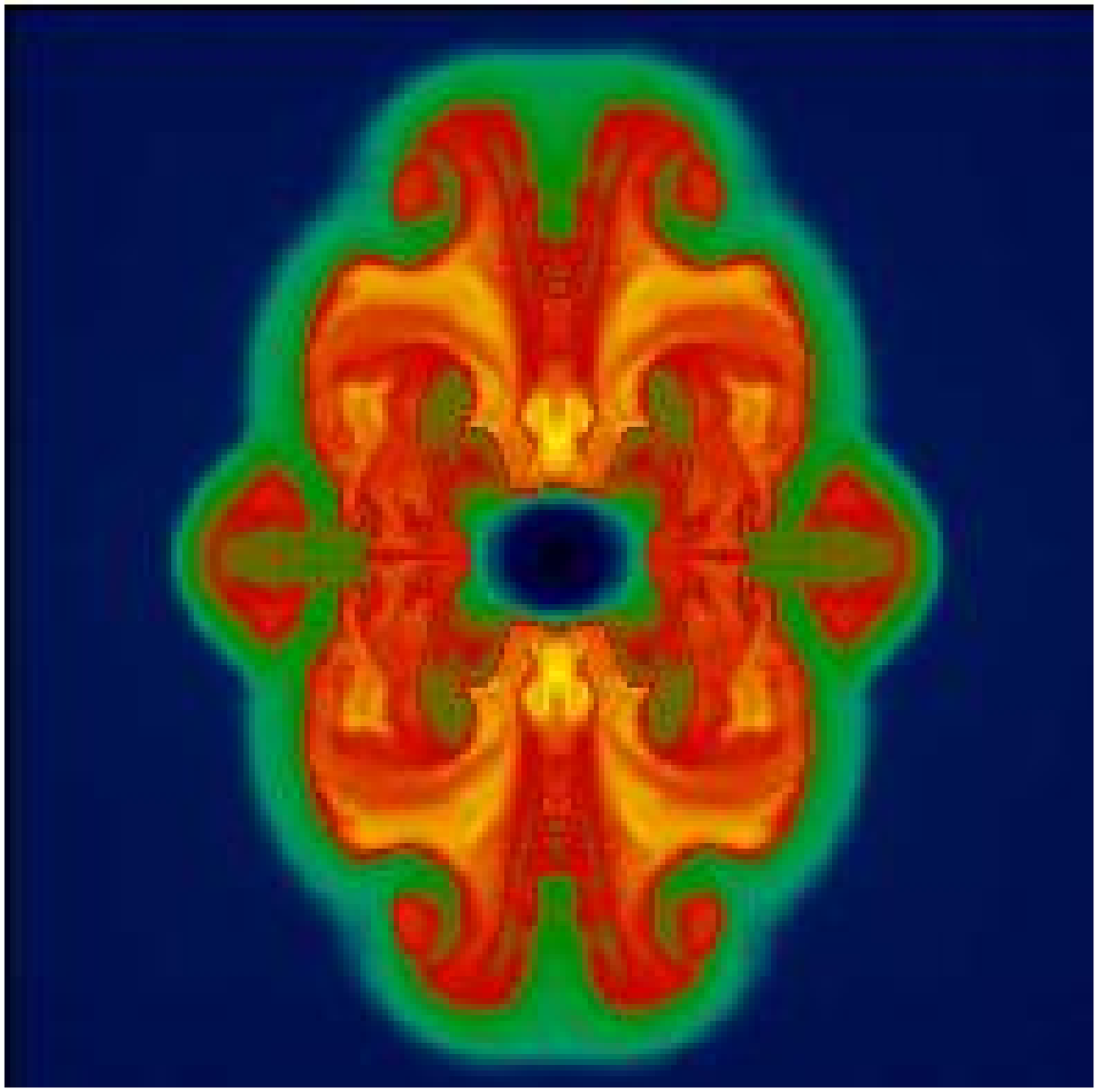}
              \epsfxsize=0.49\textwidth \epsfclipon \epsffile{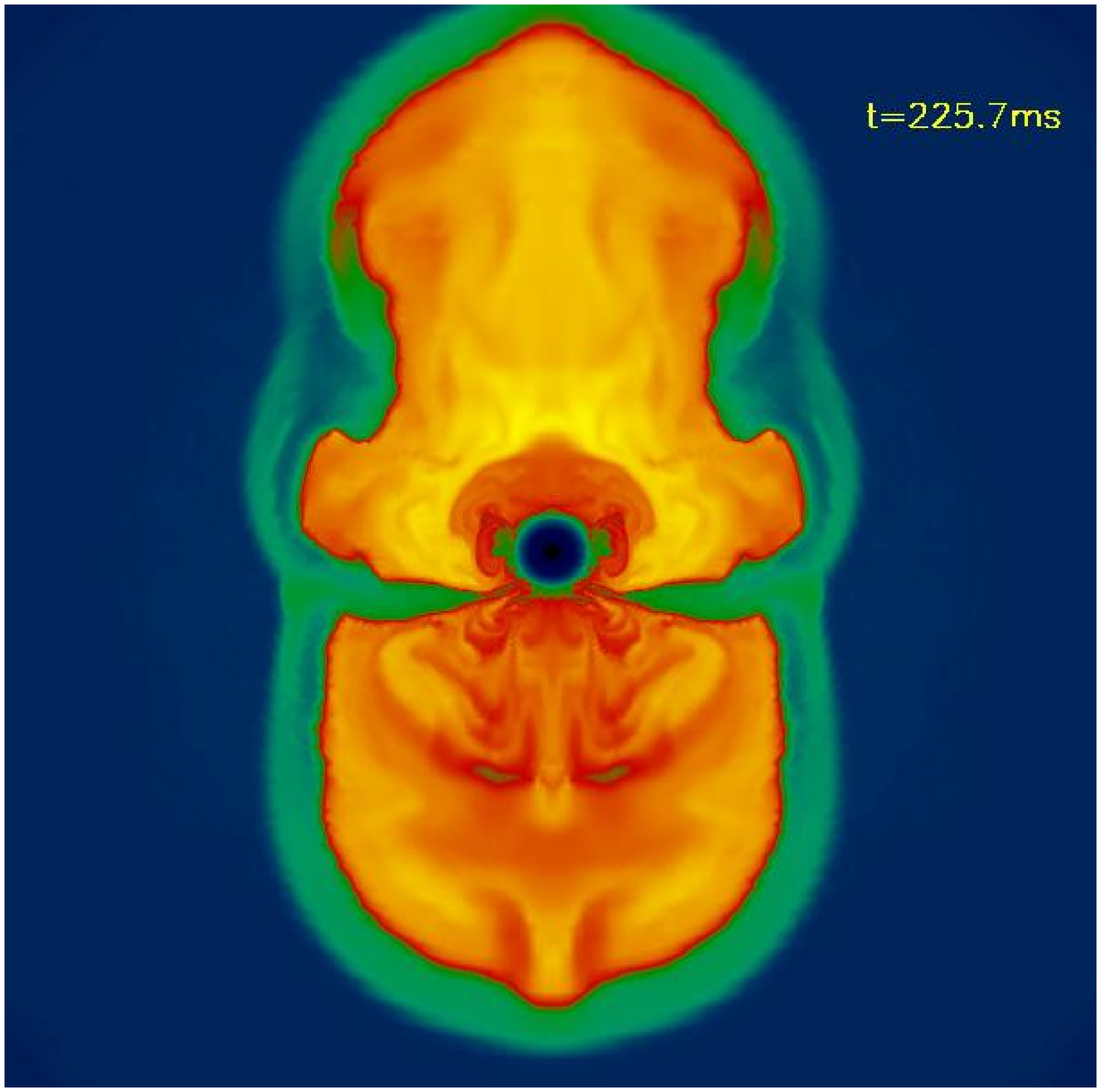}}
  \caption{\small 
Snapshots of the entropy distribution at $t > 200\,$ms post bounce for
two two-dimensional simulations with ``Boltzmann neutrino transport''
as decribed in Sect.~\ref{janka.sec_methods}. Yellow denotes highest
entropies (around 15--25$\,k_{\mathrm B}$ per nucleon), red, green,
and blue ($\la 8\,\,k_{\mathrm B}$ per nucleon) successively lower values. 
{\em Left:} Result at 245$\,$ms p.b.\ for a 15$\,M_{\odot}$ progenitor
whose Fe core was assumed to rotate rigidly with 
$\Omega = 0.5\,$rad$\,$s$^{-1}$~\cite{janka.ref5,janka.ref6}, 
which is in the ballpark of predictions
from stellar evolution models~\cite{janka.ref_heg03}. Equatorial
symmetry was assumed. At the displayed time the shock reaches its largest
expansion of nearly 300$\,$km in the axial (in the plot vertical) direction,
but lateron contracts again.
{\em Right:} Situation in an 11.2$\,M_{\odot}$ star at 226$\,$ms post bounce
(when the simulation had to be stopped because of shortage of 
computer time)~\cite{janka.ref8}.
The shock has reached a maximum radius of 600$\,$km with no sign of 
return. Global, violent low-mode bipolar oscillations of the postshock 
layer, which were possible in this simulation due to a 180$^{\mathrm o}$
grid, support a weak explosion in this {\em nonrotating} model. Note the
significant polar to equatorial deformation. 
}
  \label{janka.fig4}
\end{figure}

\subsection{Two-dimensional models}

Convective overturn in the neutrino-heated postshock layer
plays an important role during the supernova evolution and
can provide the crucial push for delayed shock 
revival by an enhancement of the efficiency of neutrino energy
transfer~\cite{janka.ref_heretal94,janka.ref_buretal95,janka.ref7}.
This is also seen in our recent simulations which employ a much
improved treatment of neutrino transport compared to previous 
models. Although we found convective enhancement of shock 
expansion~\cite{janka.ref5}, we were, however,
not able to confirm the successful explosions
obtained in simulations with considerable approximations of the 
neutrino physics~\cite{janka.ref_fryer}.

Rotation of the Fe core of the progenitor star, even at a 
moderate rate, supports strong postshock convection and 
brings the star much closer to an explosion by the 
neutrino-heating mechanism. In a simulation of a 15$\,M_{\odot}$
progenitor we assumed the Fe core to rotate rigidly with 
a rate of 
$\Omega = 0.5\,$rad$\,$s$^{-1}$~\cite{janka.ref5,janka.ref6},
which is in the ballpark of results from stellar evolution
calculations~\cite{janka.ref_heg03} and significantly slower
than adopted in other recent works (e.g., by Burrows and
collaborators~\cite{janka.ref_ott04} and 
Kotake et al.~\cite{janka.ref_kot03}, see also their 
contributions to
this meeting). Our rotating model showed shock expansion to
a maximum radius of nearly 300$\,$km along the rotation axis
(Fig.~\ref{janka.fig4}, left), whereas postshock convection
in the same stellar model, but without rotation, was not
strong enough to drive the shock much farther out
than in spherical symmetry~\cite{janka.ref5}.

All 2D simulations in Ref.~\cite{janka.ref5}, however, were
computed with a lateral wedge of about 90$^{\rm o}$. In case
of the rotating model, for example, equatorial symmetry was
assumed. This restriction, however, constrains the possible
modes of the flow pattern in the convective layer. 
The calculations in fact show growing power in large
structures at later times after bounce, and models with a 
full 180$^{\rm o}$ grid tend to develop a dominance of 
the $l = 1,\,2$ modes in connection with global dipolar
oscillations of the postshock layer, a phenomenon which is 
also seen in the three-dimensional case
(cf.\ Refs.~\cite{janka.ref_blonetal03,janka.ref9}
and the contribution by Scheck at this meeting).
A first 2D simulation with a 180$^{\rm o}$ grid and accurate 
neutrino transport for a (nonrotating) 11.2$\,M_{\odot}$ 
star reveals a dramatic difference compared to the 
90$^{\rm o}$ run of this progenitor in Ref.~\cite{janka.ref5}: 
The shock expands continuously with no sign of return until
the simulation was stopped at 226$\,$ms post bounce. At that
time the expanding shock is highly deformed and has reached
a polar radius of more than 600$\,$km (Fig.~\ref{janka.fig4},
right). In the 90$^{\rm o}$ simulation the shock was 
contracting again at that time~\cite{janka.ref5}.

\section{Conclusions}

The results reviewed above demonstrate the limits of
current state-of-the-art supernova simulations but also 
indicate the frontiers where ultimate success might be 
achieved in modeling convectively supported supernova 
explosions. Rotation of the stellar core, even at a 
moderate rate, has a strong supporting influence on shock 
expansion. The same is true for low-mode convection and 
global nonradial
shock instabilities, which are captured by 2D simulations
only in case of a full 180$^{\rm o}$ grid and sufficient
resolution to follow the amplification of perturbations 
between shock and neutron star by the ``advective-acoustic
cycle''~\cite{janka.ref_foglizzo}.
The incompletely known nuclear equation of state adds
a component of uncertainty to the problem. Our 1D simulations
with different EoSs show interesting and significant 
differences, but corresponding multi-dimensional 
simulations with accurate
neutrino transport have yet to be done.

{\small
\subsection*{Acknowledgements}
We are grateful to K.~Nomoto, A.~Heger, S.~Woosley, and 
M.~Limongi for providing us with their progenitor data.
Supercomputer time at the John von Neumann Institute for
Computing in J\"ulich and the Rechenzentrum Garching is
acknowledged. This work was supported by the 
Sonderforschungs\-be\-reich 375 ``Astroparticle Physics''
and the Sonderforschungsbereich/Transregio~7 
``Gravitational Wave Astronomy'' of the 
Deutsche Forschungsgemeinschaft.

}


\end{document}